\documentclass[preprint,prd,aps,showpacs,showkeys,nofootinbib]{revtex4}
\usepackage{graphicx}
\usepackage{dcolumn}
\usepackage{bm}
\usepackage{color}
\usepackage[dvipsnames]{xcolor}
\usepackage{amssymb}
\usepackage{epstopdf}

\definecolor{black-blue}{RGB}{77,116,175}
\definecolor{black-yellow}{RGB}{231,162,33}
\definecolor{black-green}{RGB}{144,180,58}
\definecolor{black-red}{RGB}{246,95,50}

\begin{document}

\title{Right-handed neutrino dark matter in $U(1)_X$SSM}
\author{Ming-Yue Liu$^{1,2,3}$, Shu-Min Zhao$^{1,2,3}$\footnote{zhaosm@hbu.edu.cn, 2826018216@qq.com}, Song Gao$^{1,2,3}$, Long Ruan$^{1,2,3}$, Tai-Fu Feng$^{1,2,3,4}$.}

\affiliation{$^1$ Department of Physics, Hebei University, Baoding 071002, China}
\affiliation{$^2$ Hebei Key Laboratory of High-precision Computation and Application of Quantum Field Theory, Baoding, 071002, China}
\affiliation{$^3$ Hebei Research Center of the Basic Discipline for Computational Physics, Baoding, 071002, China}
\affiliation{$^4$ Department of Physics, Chongqing University, Chongqing 401331, China}
\date{\today}

\begin{abstract}
There is strong evidence for the existence of dark matter in a number of current experiments. We study dark matter using the $U(1)_X$SSM obtained from the $U(1)_X$ extension of the minimal supersymmetric standard model (MSSM). In the $U(1)_X$SSM, we use the right-handed neutrino as a dark matter candidate, whose lightest mass eigenstate has cold dark matter features. In this paper, the relic density of right-handed neutrino as dark matter is investigated. For dark matter scattering, both spin-independent and spin-dependent cross sections are studied. In the final numerical results obtained, some parameter spaces can satisfy the constraints of the relic density and dark matter direct detection experiments.
\end{abstract}

\keywords{Supersymmetry,~dark matter,~neutrino.}

\maketitle

\section{Introduction}
Dark matter is a theoretical form of matter that neither emits nor absorbs any form of electromagnetic radiation (e.g., light), and thus cannot be directly observed by our telescopes. The existence of dark matter was confirmed in the 1970s when astronomical observation techniques improved significantly, and the rotation curves of galaxies were observed to contradict Newton's laws. Over the past 20 years, with the rapid advancement of technology, precise cosmological measurements have indicated that at least $85\%$ of the matter in the Universe is 'dark' that is, electrically neutral, non-luminous, and non-baryonic \cite{-1,-2}. Its presence is essential for explaining phenomena such as the rotation curves of galaxies \cite{-3}, gravitational lensing \cite{-4}, structure formation \cite{-5}, and the anisotropy of the Cosmic Microwave Background (CMB). Further evidence supporting the existence of dark matter can be found in Refs. \cite{-6,-7,-8,-9,-10}.

In order to maintain the relic density of dark matter, it should be stable or have a long lifetime. For many years there has been a great deal of interest in dark matter, but the nature of its mass and interactions has not yet been discovered. The density of non-baryonic matter is $\Omega h^2=0.12\pm0.0012$ \cite{2}, and the Standard Model (SM) cannot explain this. This means that there must be new physics beyond the SM.  Axions, inert neutrinos, primordial black holes and WIMPs \cite{10,11,12} are a few dark matter candidates. WIMPs, in particular, are among the most popular dark matter candidates. Since primordial black holes are not composed of elementary particles but are instead formed directly from the energy density of the early universe, they are considered a possible candidate for dark matter beyond the Standard Model. Additionally, PBHs can have a wide range of masses, from less than a mountain to several solar masses, making them an attractive candidate for dark matter. In summary, PBHs are mentioned as a candidate for dark matter because they do not belong to the realm of the Standard Model but represent a new physical phenomenon, which is why they are considered part of BSM physics. Due to the relatively weak interactions, WIMPs have an extremely low probability of interacting with conventional matter, which explains why they are difficult to detect directly, and is also consistent with the fundamental property that dark matter does not participate in electromagnetic interactions (therefore does not emit light). Its detection is essential to distinguish new physical models and to understand the nature of dark matter. The direct detection for dark matter is studying the recoil energy of nuclei caused by the elastic scattering of a WIMP off a nucleon. The interactions between WIMPs and nuclei can be either spin-dependent (SD) or spin-independent (SI), and their cross sections are restricted in Ref. \cite{13}. These two modes of interaction have important implications for the sensitivity and constraints of direct dark matter detection experiments.

Physicists extend the SM and obtain a large number of extended models,~~among which the Minimal Supersymmetric Standard Model (MSSM) is the most popular model. However,~it has been gradually found that there are also problems in the MSSM such as the $\mu$ problem \cite{14} and zero-mass neutrinos \cite{15}.~In order to solve these problems well, we carry out the $U(1)$ extension of the MSSM, and on the basis of the MSSM, we add three right-handed neutrinos and three singlet Higgs superfields, and obtain the $U(1)_X$SSM \cite{1,16,17}. Among them, the singlet Higgs superfields solve the $\mu$ problem, and the right-handed neutrino solves the zero-mass neutrino problem. The right-handed neutrino plays a key role in the process of baryon number asymmetry generation and it has attracted much attention as a candidate for dark matter \cite{18,19,20}. In our previous work \cite{1,21},~scalar neutrino dark matter and light neutrino dark matter are supposed as dark matter candidate in the framework of $U(1)_X$SSM. In this paper, right-handed neutrino is used as dark matter for the study at $U(1)_X$SSM. In Ref. \cite{22}, it is shown that not only can right-handed neutrinos be thermally produced enough to explain dark matter in the universe, but their elastic scattering cross section with the nucleus is large enough to detect them through the Higgs exchange process.

The lightest right-handed neutrino is an inert neutrino. In the $U(1)_XSSM$, the superpotential and Lagrangian are invariant under matter parity ($P_M$) transformations, then this also implies that the model has R-parity ($P_R$) conservation.
\begin{eqnarray}
&&P_M=(-1)^{3(B-L)},~~~~~~~P_R=(-1)^{3(B-L)+2s}.
\end{eqnarray}
where $B$ and $L$ represent the baryon number and lepton number, respectively, and $s$ is the spin of the particle.

This implies that the lightest R-parity odd state, i.e.,
the lightest supersymmetric particle (LSP),
tends to be stable and can contribute to the dark matter
density. The lightest right-handed neutrino is inert, and just participates in weak interaction.
Its life time is longer than the proton. So it  can be a candidate for dark matter.

In this work, we present a systematic analysis of the properties of right-handed neutrino as dark matter candidate at $U(1)_X$SSM. In Sec.II, we describe $U(1)_X$SSM in detail and list the mass matrix of the right-handed neutrino and some of the vertices we need. In Sec.III, we assume the right-handed neutrino as a dark matter candidate and study its relic density. In Sec.IV, we investigate the results of direct detection of scattering by right-handed neutrinos, including spin-independent and spin-correlated cross sections. In Sec.V, we calculate numerical results for the residual density and dark matter scattering cross sections, and we provide a brief review of the results and draw pictures. We discuss and draw conclusions in Sec.VI.

\section{The essential content of $U(1)_X$SSM}
In this section, we briefly describe the basic features of the $U(1)_X$SSM. $U(1)_X$SSM is obtained by extending the MSSM. The local gauge group is $SU(3)_C\times SU(2)_L \times U(1)_Y\times U(1)_X$ \cite{1}. In this model, new superfields are added to the MSSM, namely: three Higgs singlets $\hat{\eta},~\hat{\bar{\eta}},~\hat{S}$, along with right-handed neutrino superfields $\hat{\nu}_i$, serves several purposes: $\textcircled{1}$ Higgs singlets do not carry any standard model charges but may acquire vacuum expectation values, leading to spontaneous symmetry breaking of the $U(1)_X$ symmetry and possibly generating mass terms for other particles. $\textcircled{2}$ Right-handed neutrino superfields are essential for implementing seesaw mechanism to generate tiny but non-zero masses for left-handed neutrinos through Majorana mass terms. These right-handed neutrinos are sterile under the SM interactions, meaning they don't interact via electromagnetic, or strong forces except through gravity and possibly the new $U(1)_X$ interaction. The neutral CP-even parts of $H_{u}$, $H_{d}$, $\eta$, $\bar{\eta}$ and $S$ are mixed together, which is the source of the 5$\times$5 mass squared matrix. In order to obtain the Higgs mass of 125.25$\pm$0.17 GeV \cite{2}, loop corrections to the lightest CP-even Higgs particle are needed. These sneutrinos become CP-even sneutrinos and CP-odd sneutrinos, both of which have a mass square matrix of 6$\times$6.

The representation of the superpotential and the soft breaking terms in the $U(1)_X$SSM is:
\begin{eqnarray}
&&W=l_W\hat{S}+\mu\hat{H}_u\hat{H}_d+M_S\hat{S}\hat{S}-Y_d\hat{d}\hat{q}\hat{H}_d-Y_e\hat{e}\hat{l}\hat{H}_d+\lambda_H\hat{S}\hat{H}_u\hat{H}_d
\nonumber\\&&\hspace{0.6cm}+\lambda_C\hat{S}\hat{\eta}\hat{\bar{\eta}}+\frac{\kappa}{3}\hat{S}\hat{S}\hat{S}+Y_u\hat{u}\hat{q}\hat{H}_u+Y_X\hat{\nu}\hat{\bar{\eta}}\hat{\nu}
+Y_\nu\hat{\nu}\hat{l}\hat{H}_u.
\end{eqnarray}
\begin{eqnarray}
&&\mathcal{L}_{soft}=\mathcal{L}_{soft}^{MSSM}-B_SS^2-L_SS-\frac{T_\kappa}{3}S^3-T_{\lambda_C}S\eta\bar{\eta}
+\epsilon_{ij}T_{\lambda_H}SH_d^iH_u^j\nonumber\\&&\hspace{1cm}
-T_X^{IJ}\bar{\eta}\tilde{\nu}_R^{*I}\tilde{\nu}_R^{*J}
+\epsilon_{ij}T^{IJ}_{\nu}H_u^i\tilde{\nu}_R^{I*}\tilde{l}_j^J
-m_{\eta}^2|\eta|^2-m_{\bar{\eta}}^2|\bar{\eta}|^2-m_S^2S^2\nonumber\\&&\hspace{1cm}
-(m_{\tilde{\nu}_R}^2)^{IJ}\tilde{\nu}_R^{I*}\tilde{\nu}_R^{J}
-\frac{1}{2}\Big(M_S\lambda^2_{\tilde{X}}+2M_{BB^\prime}\lambda_{\tilde{B}}\lambda_{\tilde{X}}\Big)+h.c~.
\end{eqnarray}

The vacuum expectation values (VEVs) of the two Higgs doublet states $H_{u}, H_{d}$ are $v_u$, $v_d$ and the VEVs of the three Higgs singlet states $\eta$, $\bar{\eta}$, $S$ are $~v_\eta$, $v_{\bar\eta}$ and $v_S$ respectively. The Higgs superfields are displayed as follows:
\begin{eqnarray}
&&\hspace{1cm}H_{u}=\left(\begin{array}{c}H_{u}^+\\{1\over\sqrt{2}}\Big(v_{u}+H_{u}^0+iP_{u}^0\Big)\end{array}\right),
~~~~~~
H_{d}=\left(\begin{array}{c}{1\over\sqrt{2}}\Big(v_{d}+H_{d}^0+iP_{d}^0\Big)\\H_{d}^-\end{array}\right),
\nonumber\\&&\eta={1\over\sqrt{2}}\Big(v_{\eta}+\phi_{\eta}^0+iP_{\eta}^0\Big),~~~
\bar{\eta}={1\over\sqrt{2}}\Big(v_{\bar{\eta}}+\phi_{\bar{\eta}}^0+iP_{\bar{\eta}}^0\Big),~~
S={1\over\sqrt{2}}\Big(v_{S}+\phi_{S}^0+iP_{S}^0\Big).
\end{eqnarray}

The particle content and charge distribution of $U(1)_X$SSM are shown in Table \ref {JJ1}. Apparently, $U(1)_X$SSM is more complex than MSSM. We have shown that $U(1)_X$SSM is anomaly free, the details of which can be found in Ref. \cite{1}. In the $U(1)_X$SSM, the two Abelian groups $U(1)_Y$ and $U(1)_X$ lead to gauge kinetic mixing. This effect is a feature outside the MSSM and can also be induced by RGEs. $A_\mu^{'Y}$ and $A_\mu^{'X}$ denote the gauge fields of $U(1)_Y$ and $U(1)_X$ respectively. We write the covariant derivative of the $U(1)_X$SSM \cite{3,4,5,6}:
\begin{eqnarray}
&&D_\mu=\partial_\mu-i\left(\begin{array}{cc}Y^Y,&Y^X\end{array}\right)
\left(\begin{array}{cc}g_{Y},&g{'}_{{YX}}\\g{'}_{{XY}},&g{'}_{{X}}\end{array}\right)
\left(\begin{array}{c}A_{\mu}^{\prime Y} \\ A_{\mu}^{\prime X}\end{array}\right)\;.
\end{eqnarray}

\begin{table}
\caption{ The superfields in $U(1)_X$SSM}
\begin{tabular}{|c|c|c|c|c|c|c|c|c|c|c|c|}
\hline
Superfields & $\hspace{0.1cm}\hat{q}_i\hspace{0.1cm}$ & $\hat{u}^c_i$ & $\hspace{0.2cm}\hat{d}^c_i\hspace{0.2cm}$ & $\hat{l}_i$ & $\hspace{0.2cm}\hat{e}^c_i\hspace{0.2cm}$ & $\hat{\nu}_i$ & $\hspace{0.1cm}\hat{H}_u\hspace{0.1cm}$ & $\hat{H}_d$ & $\hspace{0.2cm}\hat{\eta}\hspace{0.2cm}$ & $\hspace{0.2cm}\hat{\bar{\eta}}\hspace{0.2cm}$ & $\hspace{0.2cm}\hat{S}\hspace{0.2cm}$ \\
\hline
$SU(3)_C$ & 3 & $\bar{3}$ & $\bar{3}$ & 1 & 1 & 1 & 1 & 1 & 1 & 1 & 1  \\
\hline
$SU(2)_L$ & 2 & 1 & 1 & 2 & 1 & 1 & 2 & 2 & 1 & 1 & 1  \\
\hline
$U(1)_Y$ & 1/6 & -2/3 & 1/3 & -1/2 & 1 & 0 & 1/2 & -1/2 & 0 & 0 & 0  \\
\hline
$U(1)_X$ & 0 & -1/2 & 1/2 & 0 & 1/2 & -1/2 & 1/2 & -1/2 & -1 & 1 & 0  \\
\hline
\end{tabular}
\label{JJ1}
\end{table}

Considering these two Abelian gauge groups to be complete, we change the basis by a correct matrix $R$ \cite{3,5,6} and redefine the $U(1)$ gauge field:
\begin{eqnarray}
&&\left(\begin{array}{cc}g_{Y},&g{'}_{YX}\\g{'}_{XY},&g{'}_{X}\end{array}\right)
R^T=\left(\begin{array}{cc}g_{1},&g_{YX}\\0,&g_{X}\end{array}\right)~,~~~~
R\left(\begin{array}{c}A_{\mu}^{\prime Y} \\ A_{\mu}^{\prime X}\end{array}\right)
=\left(\begin{array}{c}A_{\mu}^{Y} \\ A_{\mu}^{X}\end{array}\right)\;.
\end{eqnarray}

At the tree level, the three neutral gauge bosons $A_\mu^X$, $A_\mu^Y$ and $A_\mu^3$ are mixed together and their mass matrices are displayed in the basis ($A_\mu^Y$, $A_\mu^3$, $A_\mu^X$) \cite{7}, the corresponding mass matrice reads as:
\begin{eqnarray}
&&\left(\begin{array}{*{20}{c}}
\frac{1}{8}g_{1}^2 v^2 &~~~ -\frac{1}{8}g_{1}g_{2} v^2 & ~~~\frac{1}{8}g_{1}(g_{{YX}}+g_{X}) v^2 \\
-\frac{1}{8}g_{1}g_{2} v^2 &~~~ \frac{1}{8}g_{2}^2 v^2 & ~~~~-\frac{1}{8}g_{2}g_{{YX}} v^2\\
\frac{1}{8}g_{1}(g_{{YX}}+g_{X}) v^2 &~~~ -\frac{1}{8}g_{2}(g_{{YX}}+g_{X}) v^2 &~~~~ \frac{1}{8}(g_{{YX}}+g_{X})^2 v^2+\frac{1}{8}g_{{X}}^2 \xi^2
\end{array}\right),\label{gauge matrix}
\end{eqnarray}
with $v^2=v_u^2+v_d^2$ and $\xi^2=v_\eta^2+v_{\bar{\eta}}^2$.

Mass matrix for neutrinos, basis: ($\nu_L$, $\bar{\nu}_R$)
\begin{eqnarray}
m_\nu = \left(
\begin{array}{cc}
0 &\frac{1}{\sqrt{2}}v_uY_\nu^T\\
\frac{1}{\sqrt{2}}v_uY_\nu &\sqrt{2}v_{\bar{\eta}}Y_x\end{array}
\right),\label{G2}
 \end{eqnarray}
This matrix is diagonalized by $U^V$:
\begin{eqnarray}
U^{V,*}m_\nu U^{V,\dag} = m_\nu^{dia}.
 \end{eqnarray}

Other mass matrices can be found in Refs. \cite{8,9}.

We show some of the couplings needed in the $U(1)_X$SSM, and derive the vertices of $\nu_i-\nu_j-h_k$:
\begin{eqnarray}
&&\mathcal{L}_{\nu_i\nu_jh_k}=-\frac{i}{\sqrt{2}}\sum_{b=1}^3\sum_{a=1}^3\Big[\Big((U_{jb}^{V,*}U_{i3+a}^{V,*}
+U_{ib}^{V,*}U_{j3+a}^{V,*})Y_{\nu,ab}Z_{k2}^H
\nonumber\\&&\hspace{1.2cm}+(U_{j3+b}^{V,*}U_{i3+a}^{V,*}
+U_{i3+b}^{V,*}U_{j3+a}^{V,*})Y_{x,ab}Z_{k4}^H\Big)P_L
\nonumber\\&&\hspace{1.2cm}+\Big((U_{j3+a}^{V}U_{ib}^{V}Z_{k2}^H+U_{i3+a}^{V}U_{ib}^{V})Y_{\nu,ab}^*Z_{k2}^H
\nonumber\\&&\hspace{1.2cm}+(U_{j3+a}^{V}U_{i3+b}^{V}+U_{i3+a}^{V}U_{j3+b}^{V})Y_{x,ab}^*Z_{k4}^H\Big)P_R\Big].\label{nnh}
\end{eqnarray}
The vertices of $h_k-W_\mu^--W_\sigma^+$:
\begin{eqnarray}
&&\mathcal{L}_{h_k-W_\mu^--W_\sigma^+}=\frac{i}{2}g_2^2(v_dZ_{i1}^H+v_uZ_{i2}^H)g_{\sigma\mu}.
\end{eqnarray}
The vertices of $\nu_i-\nu_j-Z$:
\begin{eqnarray}
&&\mathcal{L}_{\nu_i\nu_jZ}=-\frac{i}{2}\sum_{a=1}^3 \Big[ \Big((g_1\cos\theta'_\omega\sin\theta_\omega+g_2\cos\theta_\omega\cos\theta'_\omega
-g_{YX}\sin\theta'_\omega)U_{ja}^{V,*}U_{ia}^{V}
\nonumber\\&&\hspace{2cm}+(-g_X\sin\theta'_\omega+g_{XY}\cos\theta'_\omega\sin\theta_\omega) U_{j3+a}^{V,*}U_{i3+a}^{V}\Big)\Big(\gamma_\mu\cdot P_L\Big)
\nonumber\\&&\hspace{2cm}-\Big((g_1\cos\theta'_\omega\sin\theta_\omega+g_2\cos\theta_\omega\cos\theta'_\omega-
g_{YX}\sin\theta'_\omega)U_{ia}^{V,*}U_{ja}^{V}
\nonumber\\&&\hspace{2cm}+(-g_X\sin\theta'_\omega+g_{XY}\cos\theta'_\omega\sin\theta_\omega)
U_{i3+a}^{V,*}U_{j3+a}^{V}\Big)\Big(\gamma_\mu\cdot P_R\Big) \Big].
\end{eqnarray}
The vertices of $\nu_i-\nu_j-H_k^+$:
\begin{eqnarray}
&&\mathcal{L}_{\nu_j-l_i-H_k^+}=i\sum_{b=1}^3U_{jb}^{V,*}Y_{e,ib}Z_{k1}^+P_L+
i\sum_{a=1}^3Y_{\nu,ia}^*U_{j3+a}^VZ_{k2}^+P_R.\label{AA3}
\end{eqnarray}
Other vertices can be found in Ref. \cite{1}.
\section{RELIC DENSITY}
Considering the decay at the tree level with the lightest right-handed neutrinos, we study the decays $\nu_j\rightarrow \nu_i h_0$ and $\nu_j\rightarrow l_i H^\pm$ (j = 4, 5, 6; i = 1, 2, 3).
The vertices of $\nu_i-\nu_j-h_k$ and  $\nu_j-l_i-H_k^+$  are in Eqs. (\ref{nnh}) and (\ref{AA3}).
We use the following formula to obtain the decay width
\begin{eqnarray}
&&d\Gamma=\frac{1}{32\pi^2}|\mathcal{M}|^2\frac{\overrightarrow{P}_1}{M^2}d\Omega.
\end{eqnarray}

Through the aforementioned study, we find that the decay width of the lightest
 right-handed neutrino is very tiny, and its lifetime is very long. Additionally, due to the extremely small Yukawa coupling values of neutrinos, $Y_{\nu_{ij}}~(i,j=1,2,3)$ are approximately in the range of $10^{-8}$ to $10^{-6}$.
 Therefore, based on the decay width formula, we estimate its total decay width to be on the order of $10^{-41}s$,
 which makes it sufficiently stable, potentially even more stable than a proton.

Right-handed neutrinos possess theoretical simplicity and cosmological attractiveness as dark matter candidates. In this section, we assume the lightest mass eigenstate ($\nu_1^0$) of the right-handed neutrino mass matrix Eq. (\ref{G2}) as a dark matter candidate. Any WIMP candidate must satisfy the relic density constraint, and the masses of the right-handed neutrinos can be distributed over a wide range, from the sub-electron volt level to a few thousand electron volts or even higher, which would accommodate our desired dark matter model. We denote the freeze-out temperature by $T_f$, which is usually expressed as $x_F$=$\frac{M_D}{T_f}$ for subsequent calculations, where $M_D$($M_D=M_{\nu_1^0}$) is the mass of the dark matter particle. The time evolution of the dark matter number density $n_{\nu_1^0}$ with the expansion of the universe is described by the Boltzmann equation \cite{23,24,25}, i.e.:
\begin{eqnarray}
&&\frac{dn_{\nu_1^0}}{dt}=-3Hn_{\nu_1^0}-\langle\sigma v\rangle_{SA}(n_{\nu_1^0}^2-n_{\nu_1^0eq}^2)-\langle\sigma v\rangle_{CA}(n_{\nu_1^0}n_\phi-n_{\nu_1^0eq}n_{\phi eq}).\label{G3}
\end{eqnarray}

In Eq. (\ref{G3}), $\sigma$ is the annihilation cross section of the particle, $n_{\nu_1^0eq}$ is the dark matter number density at thermal equilibrium, $v$ is the relative velocity of the annihilating particle, and $H$ is the expansion rate of the universe. For $\nu_1^0$, it will undergo self-annihilation and co-annihilation with another particle $\phi$. At temperature $T_f$ the annihilation rate of $\nu_1^0$ is approximately equal to the Hubble expansion rate, which we obtain:
\begin{eqnarray}
&&\langle\sigma v\rangle_{SA}n_{\nu_1^0}+\langle\sigma v\rangle_{CA}n_\phi\sim H(T_f).
\end{eqnarray}

With the supposition $M_\phi$ $>$ $M_{\nu_1^0}$ \cite{26}:
\begin{eqnarray}
&&n_\phi=(\frac{M_\phi}{M_{\nu_1^0}})^{3/2}Exp[(M_{\nu_1^0}-M_\phi)/T_f]n_{\nu_1^0}.
\end{eqnarray}

So it ends up being:
\begin{eqnarray}
&&[\langle\sigma v\rangle_{SA}+\langle\sigma v\rangle_{CA}(\frac{M_\phi}{M_{\nu_1^0}})^{3/2}Exp[(M_{\nu_1^0}-M_\phi)/T_f]]n_{\nu_1^0}\sim H(T_f).
\end{eqnarray}

After we calculate the self-annihilation cross section $\sigma(\nu_1^0 \nu_1^{0*} \rightarrow$ anything) and the co-annihilation cross section $\sigma(\nu_1^0 \phi \rightarrow$ anything), we can obtain the annihilation rates $\langle\sigma v\rangle_{SA}$ and $\langle\sigma v\rangle_{CA}$ in the thermal history of the Universe. Next, we use a non-relativistic approximation to $\langle\sigma v\rangle$ with $v^2$ to perform a non-relativistic approximate expansion \cite{27,28,29,30}:
\begin{eqnarray}
&&\langle\sigma v\rangle=a+bv^2+O(v^4)\approx a+6b/x_F.
\end{eqnarray}

The thermally averaged cross section time velocity $\langle\sigma_{eff}v\rangle$ is extended by Edsj\"{o} and Gunn rounds, which include co-annihilation \cite{31,32}:
\begin{eqnarray}
&&\langle\sigma_{eff}v\rangle(x_F)=\frac{\int_2^\infty K_1(\frac{a}{x})\sum_{i,j=1}^N\lambda(a^2,b_i^2,b_j^2)g_ig_j\sigma_{ij}(a)da}{4x(\sum_{i=1}^NK_2
(\frac{b_i}{x})b_i^2g_i)^2}.\label{G4}
\end{eqnarray}

In Eq. (\ref{G4}),~$\lambda(a^2,b_i^2,b_j^2)=a^4+b_i^4+b_j^4-2(a^2b_i^2+a^2b_j^2+b_i^2b_j^2)$, $a=\sqrt{s}/M_{\nu_1^0}$ and $b_i=m_i/M_{\nu_1^0}$. $g$ is the number of relativistic degrees of freedom with mass less than $T_f$.

\begin{figure}[ht]
\setlength{\unitlength}{5mm}
\centering
\includegraphics[width=3in]{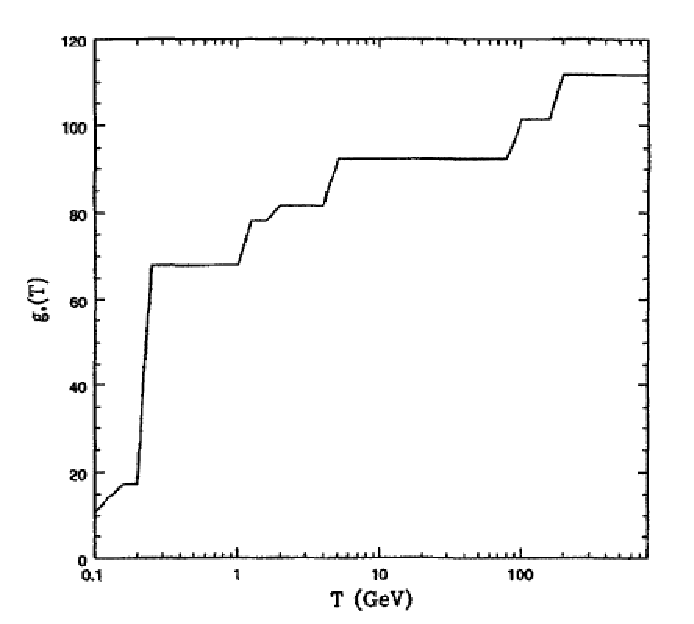}
\caption{Evolution of the degree of freedom $g_*$ with the cosmic background temperature \cite{34}.} {\label {S1}}
\end{figure}

We can obtain an iterative equation for $x_F$ which solves Eq. (\ref{G3}):
\begin{eqnarray}
&&x_F=\frac{m_{\nu_1^0}}{T_f}\simeq \log[\frac{0.076M_{pl}m_{\nu_1^0}(a+6b/x_F)}{\sqrt{g_*x_F}}],\label{GG1}
\end{eqnarray}
with $M_{pl}=1.22\times10^{19}$ GeV (Planck mass), the evolution of $g_*$ with cosmic temperature $T_f$ is shown in Fig.\ref{S1}.
We make use of the Taylor series expansion to simplify the results.
The relative velocity of the initial dark matter is slow, and is considered as small parameter to series.
In Eq.(\ref{G4}) and Eq.(\ref{GG1}), $a$
represents the expanded leading order result and $b$ represents the next leading order result.

We calculate the relic density $\Omega_Dh^2$ of cold non-baryonic dark matter and reduce it to the following form \cite{33,34}:
\begin{eqnarray}
&&\Omega_Dh^2\simeq\frac{1.07\times10^9x_F}{\sqrt{g_*}M_{pl}(a+3b/x_F)~{\rm GeV}}.
\end{eqnarray}

\begin{figure}[ht]
\setlength{\unitlength}{5.0mm}
\centering
\includegraphics[width=6.5in]{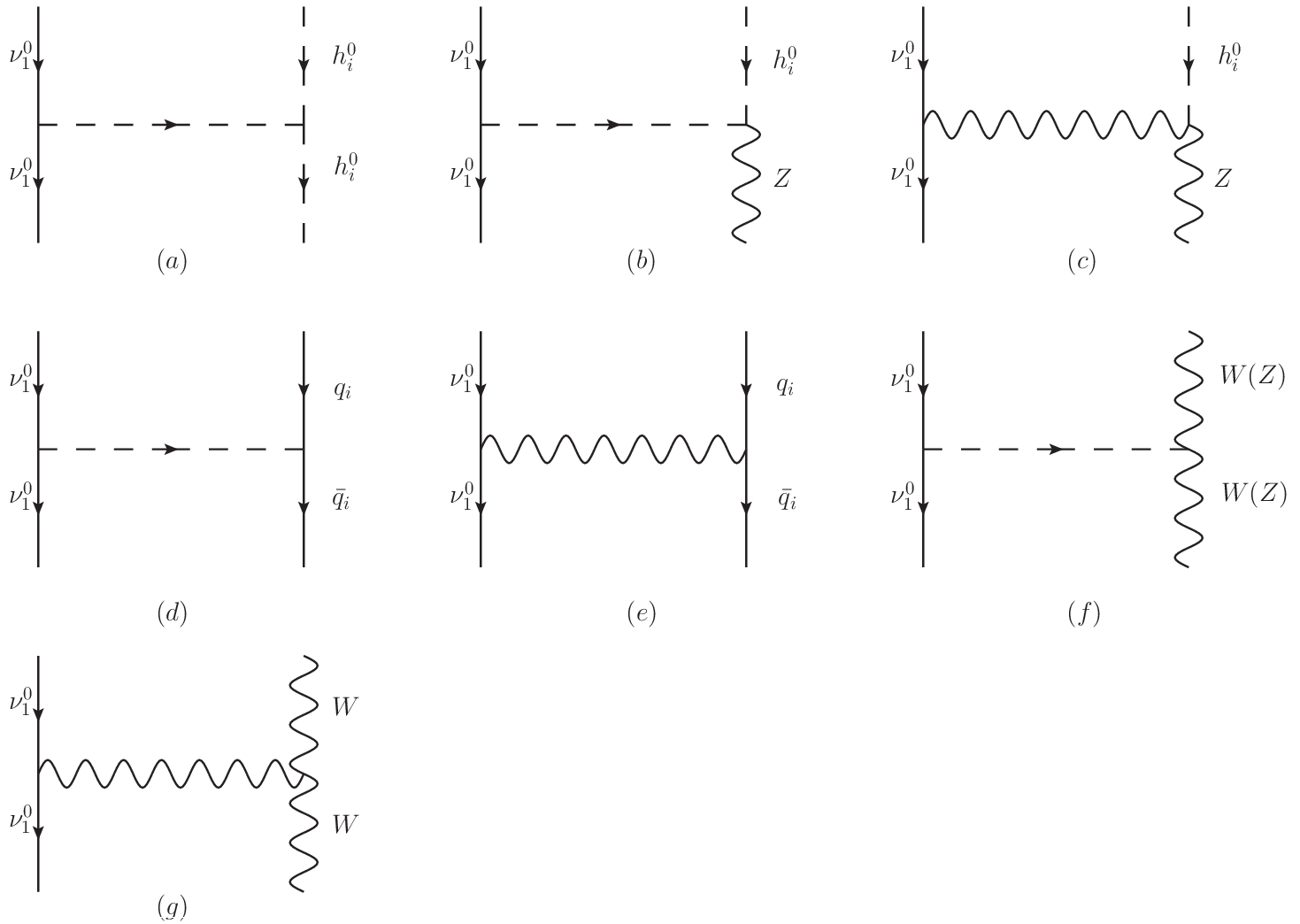}
\caption{Feynman diagrams for self-annihilation processes.}\label{A1}
\end{figure}

We perform calculations for both self-annihilation and co-annihilation, where the self-annihilation process dominates in general, ignoring some lower contributions. The topology of the self-annihilation process is plotted in Fig.\ref{A1} (there are also some plots where the contributions are so small as to be negligible, and so are not plotted here).

\section{DIRECT DETECTION}
The right-handed neutrino can easily satisfy the constraints of the direct detection experiment because it is inactive. We convert the direct detection from quark-level interactions to inter-nucleon interactions and subsequently calculate the scattering cross section of the right-handed neutrino with the nucleon. The direct detection proceeds as $\nu_{1}^0+q\rightarrow\nu_{1}^0+q$, where the exchange of $Z$ boson is largely suppressed by the neutrino Yukawa coupling and is therefore completely negligible. We use $N$ for proton or neutron and calculate as follows \cite{35}:
\begin{eqnarray}
&&a_qm_q\bar{q}q\rightarrow f_Nm_N\bar{N}N.
\end{eqnarray}

The nucleon matrix elements are usually defined:
\begin{eqnarray}
&&\langle N|m_q\bar{q}q|N\rangle=m_Nf_{T_q}^{(N)}.
\end{eqnarray}
$f_N$ contains the coupling to gluons induced by integrating over the heavy quark loops. It is calculated as:
\begin{eqnarray}
&&f_N=\sum_{q=u,d,s}f_{T_q}^{(N)}a_q+\frac{2}{27}f_{T_G}^{(N)}\sum_{q=c,b,t}a_q,
~~~f_{T_G}^{(N)}=1-\sum_{q=u,d,s}f_{T_q}^{(N)}.
\end{eqnarray}
Currently, most of the values of $f_{T_q}^{(N)}$ are those found in the DarkSUSY, and we take specific values for the parameter $f_{T_q}^{(N)}$ \cite{24,36,37} in this paper:
\begin{eqnarray}
&&f_{T_u}^{(p)}=0.0153,~~~f_{T_d}^{(p)}=0.0191,~~~f_{T_s}^{(p)}=0.0447,
\nonumber\\&&f_{T_u}^{(n)}=0.0110,~~~f_{T_d}^{(n)}=0.0273,~~~f_{T_s}^{(n)}=0.0447.
\end{eqnarray}

In the following equation, we use $Z$ to denote the number of protons, and $A$ to denote the number of atoms. For $\bar{\nu}_1^0\nu_1^0\bar{q}q$ (spin-independent operator), the scattering cross section is \cite{35}:
\begin{eqnarray}
&&\sigma=\frac{1}{\pi}\hat{\mu}^2[Zf_p+(A-Z)f_n]^2.
\end{eqnarray}

The angular momentum number of the atomic nucleus is denoted by $J_N$, and the scattering cross section for $\bar{\nu}_1^0\gamma_\mu\gamma_5\nu_1^0\bar{q}\gamma^\mu\gamma_5q$ (the spin-dependent operator) is \cite{35}:
\begin{eqnarray}
&&\sigma=\frac{16}{\pi}\hat{\mu}^2a_N^2J_N(J_N+1).
\end{eqnarray}
We list the formula corresponding to a nucleus reading as:
\begin{eqnarray}
&&\sigma=\frac{12}{\pi}\hat{\mu}^2a_N^2.
\end{eqnarray}
Here, $\hat{\mu}$ represents the effective mass of the system, $\hat{\mu}=\frac{\mu M_D}{\mu+M_D}$.
Here, $\mu$ represents the mass of nucleon (neutron or proton).
$a_q$ denotes a coefficient of a operator at the quark level.
It interacts with the nucleon and is transformed into coefficients and operators at the nucleon level.
$a_N$ stands for the coefficient at the nucleon level.

\section{Numerical analysis}
The right-handed neutrino can easily satisfy the constraints of direct detection experiments by expressing this property of the neutrino in terms of its neutrino Yukawa couplings in the mass matrix. This leads to the suppression factor $Z_\nu^{Ij}$ appearing in the Feynman rule coupling $Z-\nu_i-\nu_j$ and $A_0-\nu_i-\nu_j$. $Z_\nu^{Ij}$ is proportional to the tiny neutrino Yukawa couplings $Y_\nu$. The order of magnitude of $Y_\nu$ is in the $10^{-6}-10^{-9}$ range.

In order to ensure the rigour of our results, we restrict the parameters used by considering experimental results from the Particle Data Group (PDG) \cite{2}. In this work, the mass of the lightest neutral CP-even Higgs particle $h$ would require $m_h$=125.25$\pm$0.17 GeV.

The parameters used are as follows:
\begin{eqnarray}
&&\tan\beta=15,~~Y_{X11}=0.2,~~Y_{X22}=0.4,~~Y_{X33}=0.35,~~M_{BL}=0.1~{\rm TeV},
\nonumber\\&&T_{\lambda_H}=T_{\lambda_C}=0.1~{\rm TeV},~~l_W=B_\mu=B_S=0.5~{\rm TeV}^2,~~\lambda_C=-0.2,
\nonumber\\&&\kappa=0.06,~~M_S=2.7~{\rm TeV},~~M_1=M_2=1~{\rm TeV},M_{BB'}=0.4~{\rm TeV}~~.
\end{eqnarray}

In addition to the fixed parameters shown above, there are a number of adjustable parameters as follows:
\begin{eqnarray}
&&g_X,~~g_{YX},~~\kappa,~~v_S,~~\lambda_H,~~\mu.
\end{eqnarray}
When $g_X=0.385$, $g_{YX}=0.04$, $v_S=4.3$~TeV, $\lambda_H=0.1$, $\mu=1$~TeV, $\kappa=0.06$ (in subsequent numerical analyses, the values listed here are used, unless otherwise specified) and all non-diagonal elements of mass matrix are zero, the numerical value of the dark matter relic density turns out to be $\Omega h^2$ = 0.12047. The lightest right-handed neutrino at this time is 285.97 GeV, and $\Omega h^2$ is very close to the experimental centre value, with a sensitivity within 1$\sigma$.

\subsection{ Relic density of right-handed neutrino dark matter}
\begin{figure}[ht]
\setlength{\unitlength}{5mm}
\centering
\includegraphics[width=3in]{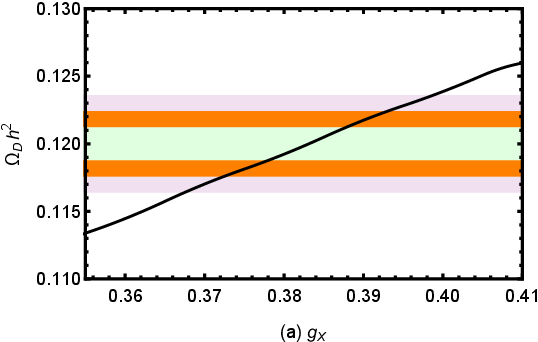}
\setlength{\unitlength}{5mm}
\centering
\includegraphics[width=3in]{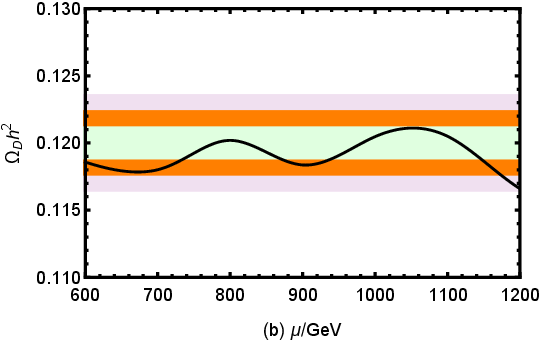}
\setlength{\unitlength}{5mm}
\centering
\includegraphics[width=3in]{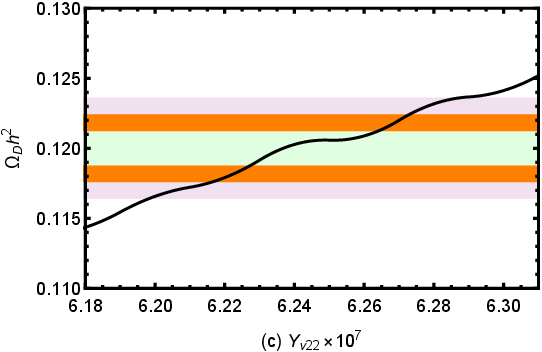}
\setlength{\unitlength}{5mm}
\centering
\includegraphics[width=3in]{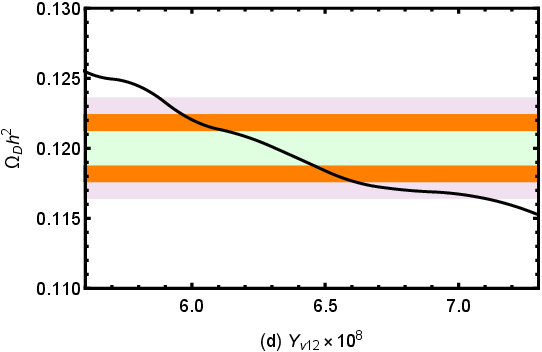}
\caption{Variation curves of relic density ($\Omega h^2$) with $g_X$, $\mu$, $Y_{\nu22}$,  $Y_{\nu12}$ in the 3$\sigma$ range.}{\label {T1}}
\end{figure}

In this section, we study the effects of different parameters on the relic density and analyse the reasons for this. In Fig. \ref{T1}, $\kappa=0.06$, $g_{YX}=0.04$, $v_S=4.3$~TeV , $\lambda_H=0.1$ and we analyse the effects of parameters $g_X$, $\mu$, $Y_{\nu22}$, $Y_{\nu12}$. Coloured regions are experimental results for $\Omega h^2$ in 3$\sigma$ (1$\sigma$ (LightGreen): 0.1188 - 0.1212; 2$\sigma$ (Orange): 0.1176 - 0.1224; 3$\sigma$ (LightPurple): 0.1164 - 0.1236).

Fig. \ref{T1} (a) shows the variation curve of $\Omega h^2$ versus $g_X$. $\Omega h^2$ increases with $g_X$ increases, and $\Omega h^2$ is within 3$\sigma$ when 0.368 $<$ $g_X$ $<$ 0.4. $g_X$ is the gauge constant, and the interaction strength it determines affects the interaction rate of right-handed neutrinos with other particles, and thus their freeze-out temperature in the early universe and subsequent dark matter relic density.  $\mu$ parameter affects the Higgs in the $U(1)_X$SSM as well as other relevant interaction strengths, which in turn affects the right-handed neutrino production and annihilation processes. So we plot the variation curve of the effect of $\mu$ on $\Omega h^2$ in Fig. \ref{T1} (b), and the results show that the relic densities can fit the experimental values very well in the range of $\mu$ (600 GeV-1200 GeV). Obviously most of the experimental values reach a sensitivity within 2$\sigma$, which means that the parameter $\mu$ has a relatively large influence on the results and is a sensitive parameter. Prior to this we fit the neutrino values, as described in Ref. \cite{8}, with $Y_{\nu22}$ on the order of $10^{-7}$ and $Y_{\nu12}$ on the order of $10^{-8}$. In Fig. \ref{T1} (c)(d), we plot an image of the Yukawa coupling coefficients $Y_{\nu22}$, $Y_{\nu12}$ versus $\Omega h^2$. It is not difficult to find that as $Y_{\nu22}$ increases $\Omega h^2$ also becomes progressively larger. However the opposite is true for $Y_{\nu12}$, which decreases as $Y_{\nu12}$ becomes larger. Both of them affect the relic density in a more regular way. Yukawa coupling is commonly used to describe the interaction between canonical bosons and fermions. However, the specific application of the Yukawa coupling in contexts involving right-handed neutrinos as dark matter candidates is not straightforward. This is because the traditional Yukawa potential describes the short-range nuclear force, not the long-range force directly associated with leptonic or neutrino dark matter. The $U(1)_X$SSM introduces new scalar fields that couple to right-handed neutrino through Yukawa interactions, which have an impact on the relic density of right-handed neutrinos when it is supposed as dark matter.

\begin{table*}
\caption{Scanning parameters for Fig.\ref{T2} and Fig.\ref{T4}.}
\begin{tabular*}{\textwidth}{@{\extracolsep{\fill}}|l|l|l|l|l|l|l|@{}}
\hline
Parameters&$~~~~~\mu$~&$g_X~$~~&$g_{YX}~~~~~$&~$\lambda_H~~~~~$&~$Y_{\nu12}$&~$Y_{\nu22}$\\
\hline
~Min&~500~GeV~~~~~&~0.2~~~~~~&0.01~~&0.05~~~~&$5\times10^{-8}~~~~$&$6\times10^{-7}~~~~~$\\
\hline
~Max&~2000~GeV~~~~~&~0.4~~~~~~&0.2~~&0.3~~~&$7\times10^{-8}~~~~$&$7\times10^{-7}~~~~~$\\
\hline
\end{tabular*}
\label{Z1}
\end{table*}

\begin{figure}[ht]
\setlength{\unitlength}{5mm}
\centering
\includegraphics[width=3in]{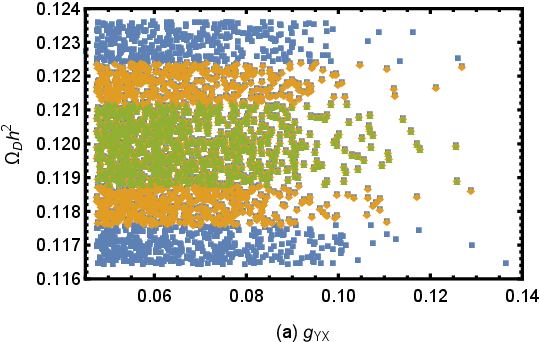}
\setlength{\unitlength}{5mm}
\centering
\includegraphics[width=3in]{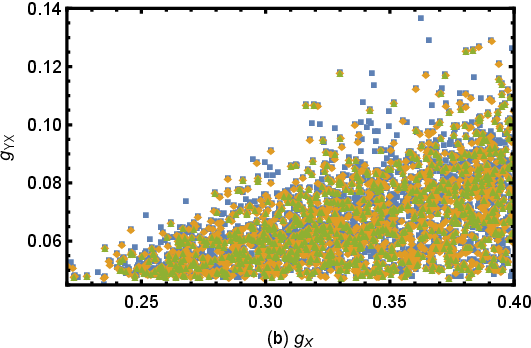}
\setlength{\unitlength}{5mm}
\centering
\includegraphics[width=3in]{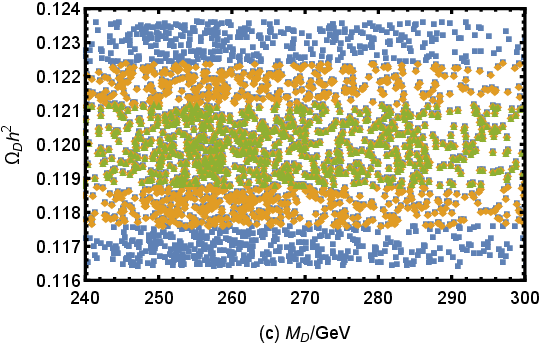}
\setlength{\unitlength}{5mm}
\centering
\includegraphics[width=3in]{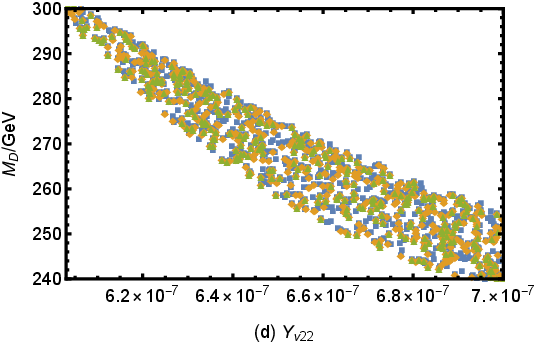}
\caption{the notation \textcolor{black-green}{$\blacktriangle$} (0.1188$\leq\Omega h^2\leq$0.1212), \textcolor{black-yellow}{$\blacklozenge$} (0.1176$\leq\Omega h^2\leq$0.1224), \textcolor{black-blue}{$\blacksquare$} (0.1164$\leq\Omega h^2\leq$0.1236).}{\label {T2}}
\end{figure}

 In this section, we scatter the points with different shapes (\textcolor{black-green}{$\blacktriangle$}$\sim1\sigma$ (0.1188$\leq\Omega h^2\leq$0.1212), \textcolor{black-yellow}{$\blacklozenge$}$\sim2\sigma$ (0.1176$\leq\Omega h^2\leq$0.1224), \textcolor{black-blue}{$\blacksquare$}$\sim3\sigma$ (0.1164$\leq\Omega h^2\leq$0.1236)) in three different colours. The range of randomised scatters is shown in Table \ref {Z1}. Fig. \ref{T2} (a) represents $\Omega h^2$ in the 3$\sigma$ range with respect to $g_{YX}$. In the figure it is clearly shown that the dots are denser for 0.04$\leq g_{YX}\leq$0.1 and very sparse when $g_{YX}>$0.1. $g_{YX}$ affects the strength of the interactions of the right-handed neutrinos with the $Z$ boson and other particles interacting via $U(1)_X$. Stronger couplings lead to greater annihilation rates, which reduces the number of neutrinos left over from the cooling process of the universe and lowers their final relic density, so producing the phenomenon shown. The relationship between $g_{YX}$ and $g_X$ is analysed as shown in Fig. \ref{T2} (b), with most of the points clustered in the lower right corner. Since we follow the 3$\sigma$ division, ~\textcolor{black-green}{$\blacktriangle$}, ~\textcolor{black-yellow}{$\blacklozenge$}, ~\textcolor{black-blue}{$\blacksquare$} are triple-stacked, which means that the range of $g_{YX}$ and $g_X$ satisfying the experimental limits on relic density is 0.2$\leq g_X\leq0.4$, 0.04$\leq g_{YX}\leq$0.11. In order to find a large parameter space that satisfies the relic density, we plot the relic density versus the lightest right-handed neutrino mass $m_D$, and the results are shown in Fig. \ref{T2} (c). In the neighbourhood of $m_D$ (200 GeV-300 GeV), we can easily see a reasonable parameter space. Fig. \ref{T2} (d) is an interesting plot, here representing the relationship between $Y_{\nu22}$ and $m_D$, still categorised according to the 3$\sigma$ classification of relic density. All points are clearly concentrated in the diagonal region and are also covered by 1$\sigma$, 2$\sigma$ and 3$\sigma$ layers.

\subsection{ The cross section of the neutrino scattering off nucleon}
In this subsection, the spin-independent and spin-dependent cross sections for the lightest neutralino scattering are investigated numerically with the parameters $\kappa$,~$v_S$, ~$g_{YX}$,~$g_{X}$,~$\mu$ and $\lambda_H$ by letting $Y_{\nu12}=6.28039\times10^{-8}$, $Y_{\nu22}=6.2483\times10^{-7}$. Taking into account constraints from the relic density, the mass of the lightest right-handed neutrino in the used parameter space is about 286 GeV. The experimental limit for direct detection of the spin-independent cross section is about $2.3\times10^{-46}~{\rm cm^2}$ within 1$\sigma$ sensitivity. The experimental constraints on the spin-independent cross section are much stricter than those on the spin-dependent cross section. The experimental limit from direct detection of spin-dependent cross sections from the Xenon1T experiment is about $4.0\times10^{-41} ~{\rm cm^2}$.

\begin{figure}[ht]
\setlength{\unitlength}{5mm}
\centering
\includegraphics[width=3in]{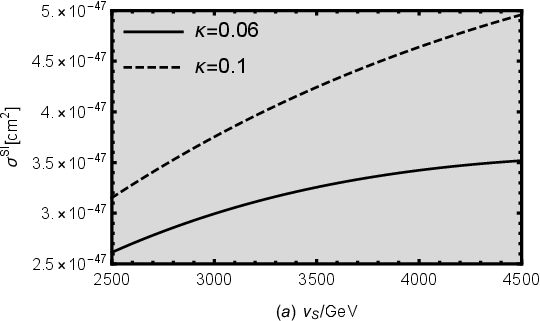}
\setlength{\unitlength}{5mm}
\centering
\includegraphics[width=3in]{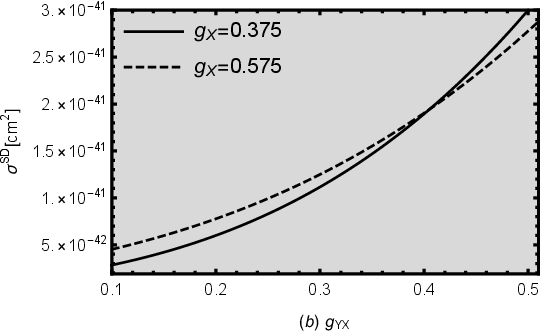}
\caption{Variation curves of cross section ($\sigma^{SI}$ and $\sigma^{SD}$) with $g_X$, $\kappa$, $v_S$,  $g_{YX}$.}{\label {T3}}
\end{figure}

In Fig. \ref{T3} (a), we discuss the spin-independent cross section, and plot $v_S$ versus $\sigma^{SI}$, where the solid line denotes $\kappa=0.06$, the dashed line denotes $\kappa=0.1$, and the grey area is the region that satisfies the experimental limits. We find that $\sigma^{SI}$ becomes larger as $v_S$ and $\kappa$ become larger. $v_S$ is the vacuum expectation value (VEV) of the Higgs singlets, which increases or decreases the strength of interactions of right-handed neutrinos with other particles. These interactions determine the probability of inelastic scattering of dark matter particles with nuclei within the detector material. A larger cross-section means that dark matter particles are more likely to collide with the detector's inner core, thus increasing the likelihood of detecting a dark matter signal. The results of our discussion on spin-dependent cross section are presented in Fig. \ref{T3} (b), which plots the curve of $\sigma^{SD}$ versus $g_{YX}$, the solid and dashed lines indicate $g_{X}$=0.375, $g_{X}$=0.575, respectively. The grey area is the area that satisfies the experimental limits. From Fig. \ref{T3} (b) it is obtained that as $g_{YX}$ increases, $\sigma^{SD}$ shows an increasing trend, in the range of $g_{YX}$(0.1-0.43). $g_{YX}$ and $\sigma^{SD}$ are positively correlated, and they are negatively correlated for $g_{YX}>$0.43.

\begin{figure}[ht]
\setlength{\unitlength}{5mm}
\centering
\includegraphics[width=3in]{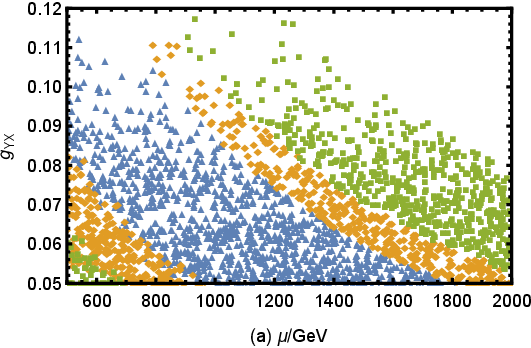}
\setlength{\unitlength}{5mm}
\centering
\includegraphics[width=3in]{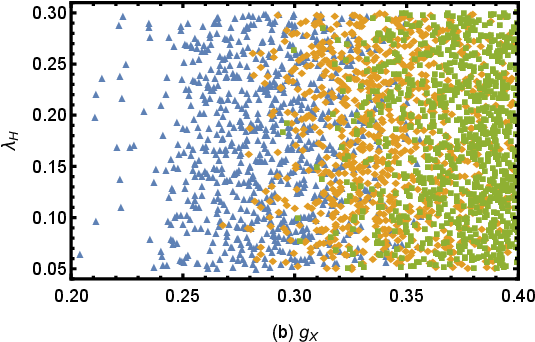}
\caption{Fig\ref{T4} (a), the notation \textcolor{black-blue}{$\blacktriangle$} ($\sigma^{SI}[{\rm cm^2}]<1.5\times10^{-47}$),~\textcolor{black-yellow}{$\blacklozenge$} ($1.5\times10^{-47}\leq\sigma^{SI}[{\rm cm^2}]<4.3\times10^{-47}$),~\textcolor{black-green}{$\blacksquare$} ($4.3\times10^{-47}\leq\sigma^{SI}[{\rm cm^2}]<2.3\times10^{-46}$); Fig\ref{T4} (b),~ \textcolor{black-blue}{$\blacktriangle$} ($\sigma^{SD}[{\rm cm^2}]<1.7\times10^{-42}$),~\textcolor{black-yellow}{$\blacklozenge$} ($1.7\times10^{-42}\leq\sigma^{SD}[{\rm cm^2}]<2.2\times10^{-42}$),~ \textcolor{black-green}{$\blacksquare$} ($2.2\times10^{-42}\leq\sigma^{SD}[{\rm cm^2}]<4\times10^{-41}$). }{\label {T4}}
\end{figure}

For Fig. \ref{T4}, scatter plots are plotted for the spin-independent and spin-dependent cross sections. Fig. \ref{T4} (a) represents the relationship between $\mu$ and $g_{YX}$, with the notation $\textcolor{black-blue}{\blacktriangle}$ ($\sigma^{SI}[{\rm cm^2}]<1.5\times10^{-47}$),~$\textcolor{black-yellow}{\blacklozenge}$ ($1.5\times10^{-47}\leq\sigma^{SI}[{\rm cm^2}]<4.3\times10^{-47}$), ~$\textcolor{black-green}{\blacksquare}$ ($4.3\times10^{-47}\leq\sigma^{SI}[{\rm cm^2}]<2.3\times10^{-46}$). The value of $\sigma^{SI}$ gradually increases from inside to outside, reaching a maximum at $\textcolor{black-green}{\blacksquare}$. Fig. \ref{T4} (b) shows the relationship between $g_X$ and $\lambda_H$, $\textcolor{black-blue}{\blacktriangle}$ ($\sigma^{SD}[{\rm cm^2}]<1.7\times10^{-42}$),~$\textcolor{black-yellow}{\blacklozenge}$ ($1.7\times10^{-42}\leq\sigma^{SD}[{\rm cm^2}]<2.2\times10^{-42}$),~ $\textcolor{black-green}{\blacksquare}$ ($2.2\times10^{-42}\leq\sigma^{SD}[{\rm cm^2}]<4\times10^{-41}$, the points are denser in the range of 0.25$\leq g_X\leq$0.4, and gradually become larger from left to right.

\section{Conclusion}
The SM has never been able to give a reasonably convincing explanation of dark matter-related phenomenology from the perspective of particle physics. Therefore, the dark matter problem becomes another strong evidence for the existence of new physics, and the development of new physics models extending the SM is an effective way to solve the dark matter problem. We extend the MSSM with the $U(1)_X$ local gauge group to obtain the so-called $U(1)_X$SSM. Comparing with MSSM, $U(1)_X$SSM has more superfields including: right-handed neutrinos and three Higgs singlets. Among the many ways to extend the SM, this paper provides a careful and in-depth analysis of the dark matter problem in the $U(1)_X$SSM.

Assuming that the lightest right-handed neutrino can be a candidate for cold dark matter, firstly, we calculate its relic density in a suitable parameter space where $\Omega h^2$ can fit the experimental limits. Next, we calculate the numerical results using the programme to find the sensitive parameters and draw a lot of these sensitive parameters against $\Omega h^2$. We find that the parameters $g_X$, $g_{YX}$, $v_S$, $\lambda_H$, $\mu$, $\kappa$, $Y_{\nu12}$ and $Y_{\nu22}$ are the sensitive parameters, which have a large impact on the results. Secondly, the scattering cross sections of dark matter and nucleons are calculated and plotted as images, separately for spin-dependent and spin-independent cross sections. The parameter space we used during the study is reasonable and satisfies the residual density and nucleon scattering constraints of dark matter. We believe that the research results in this paper can provide a possibility to explain the dark matter phenomenon, and also provide a direction for the experimental detection of dark matter, and contribute to the exploration of new physics.

\begin{acknowledgments}
This work is supported by National Natural Science Foundation of China (NNSFC)(No.12075074),
Natural Science Foundation of Hebei Province(A2020201002, A2023201040, A2022201022, A2022201017, A2023201041),
Natural Science Foundation of Hebei Education Department (QN2022173), the Project of the China Scholarship Council (CSC) (No. 202408130113),
Post-graduate's Innovation Fund Project of Hebei University (HBU2024SS042), the youth top-notch talent support program of the Hebei Province.
\end{acknowledgments}


\begin{thebibliography}{50}
\vspace{3mm}


\bibitem{-1}Planck Collaboration (P.A.R. Ade et al.). $Planck~2013~results.~XXXI.~Consistency~of~the
\nonumber\\~Planck~data[J].~Astron.~Astrophys$. 571: 1054 (2014).
\bibitem{-2}R.H. Cyburt. $Primordial~nucleosynthesis~for~the~new~cosmology:~Determining~uncertaint
\nonumber\\-ies~and~examining~concordance[J]$. Phys. Rev. D, 70: 023505 (2014).
\bibitem{-3}E. Corbelli, P. Salucci. $The~Extended~Rotation~Curve~and~the~Dark~Matter~Halo~of~M33[J].
\nonumber\\~Mon.Not.Roy.Astron.Soc$. 311: 441-447 (2000).
\bibitem{-4}A.N.Taylor, S.Dye, T. J. Broadhurst, N. Benitez, E. Kampen. $Gravitational~lens~magnificat
\nonumber\\-ion~and~the~mass~of~abell~1689[J].~Astrophys.J$. 501: 539 (1998).
\bibitem{-5}V. Springel, S.D.White, A.Jenkins et al. $Simulating~the~joint~evolution~of~quasars,~galaxies
\nonumber\\~and~their~large-scale~distribution[J]$. Nature, 435: 629 (2005).
\bibitem{-6}Clowe D, Bradac M, Gonzalez A H, et al. Astrophys.J., 648: 109-113 (2006).
\bibitem{-7}Taylor A, Dye S, Broadhurst T J, et al. Astrophys.J., 501: (1998).
\bibitem{-8}D. Walsh, R.F. Carswell, R.J. Weymann. Nature., 279: 381-384 (1979).
\bibitem{-9}Marusa Bradac, Steven W. Allen, et al. Astrophys.J., 687: (2008).
\bibitem{-10}S. Penny, C. Conselice, et al. Mon.Not.Roy.Astron., 393: 1054-1062 (2009).
\bibitem{2}R. L. Workman et al (PDG), Prog. Theor. Exp. Phys. 2022, 083C01 (2022).
\bibitem{10}M. Drees, M.M. Nojiri, $The~Neutralino~relic~density~in~minimal~N=1~supergravity$, Phys. Rev.D {\bf 47} 376 (1993).
\bibitem{11}S. Andreas, T. Hambye, $WIMP~dark~matter,~Higgs~exchange~and~DAMA$, JCAP {\bf 0810} 034  (2008).
\bibitem{12}J.J. Cao, Z.X. Heng, J.M. Yang, et al., $Higgs~decay~to~dark~matter~in~low~energy~SUSY:~is~it
\nonumber\\~detectable~at~the~LHC$, JHEP {\bf 1206} 145 (2012).
\bibitem{13}D, Bauer et al.,$Snowmass~CF1~Summary:~WIMP~Dark~Matter~Direct~Detection$.
\bibitem{14}U. Ellwanger, C. Hugonie, A.M. Teixeira, Phys. Rep. \textbf{496}, 1-77 (2010).
\bibitem{15}B. Yan, S.M. Zhao, T.F. Feng, Nucl. Phys. B. {\bf 975}, 115671 (2022).
\bibitem{1}S.M. Zhao, T.F. Feng, M.J. Zhang, et al., $Scalar~neutrino~dark~matter~in~U(1)_XSSM$, J. High Energy Phys. {\bf 02} 130 (2020).
\bibitem{16}F. Staub, Comput. $SARAH~4:~A~tool~for~(not~only~SUSY)~model~builders$, Phys. Commun. {\bf 185} 1773 (2014).
\bibitem{17}F. Staub, Adv. $Exploring~new~models~in~all~detail~with~SARAH$, High Energy Phys. {\bf 2015} 840780 (2015).
\bibitem{18}D.~Suematsu,$Right-handed~neutrino~as~a~common~mother~of~baryon~number~asymmetry~and
\nonumber\\~dark~matter$, (2024) [arXiv:2402.10561 [hep-ph]].
\bibitem{19}D.~G.~Cerdeno and O.~Seto, $Right-handed~sneutrino~dark~matter~in~the~NMSSM$, JCAP \textbf{08} 032 (2009).
\bibitem{20}T.~Li and W.~Chao, $Neutrino~Masses,~Dark~Matter~and~B-L~Symmetry~at~the~LHC$, Nucl. Phys. B \textbf{843} (2011).
\bibitem{21}S.M.~Zhao, G.~Z.~Ning, J.~J.~Feng, H.~B.~Zhang, T.~F.~Feng and X.~X.~Dong, $Light~neutralino~dark~matter~in~U(1)_XSSM$, Nucl. Phys. B \textbf{969}, 115469 (2021).
\bibitem{22}D. G. Cerdeno, C. Munoz and O. Seto, $Right-handed~sneutrino~as~thermal~dark~matter$, Phys. Rev. D, {\bf 79}, 023510 (2009).
\bibitem{3}G. Belanger, J. Da Silva and H.M. Tran, $Dark~matter~in~U(1)~extensions~of~the~MSSM~with
\nonumber\\~gauge~kinetic~mixing$, Phys. Rev. D {\bf 95} 115017 (2017).
\bibitem{4}V. Barger, P. Fileviez Perez and S. Spinner, $Minimal~gauged~U(1)_{B-L}~model~with~spontaneous
\nonumber\\~R-parity~violation$, Phys. Rev. Lett. {\bf 102} 181802 (2009).
\bibitem{5}P.H. Chankowski, S. Pokorski and J. Wagner, $Z-prime~and~the~Appelquist-Carrazzone~decoupling$, Eur. Phys. J. C {\bf 47} 187 (2006).
\bibitem{6}J.L. Yang, T.F. Feng, S.M. Zhao, et al., $Two~loop~electroweak~ corrections~to~\bar{B}~\rightarrow X_s\gamma~and~B_s^0\rightarrow \mu^+\mu^-~in~the~B-LSSM$, Eur. Phys. J. C {\bf 78} 714 (2018).
\bibitem{7}S.M. Zhao, L.H. Su, X.X. Dong, T.T. Wang and T.F. Feng, $Study~muon~g-2~at
~two-loop~level~in~the~U(1)_XSSM$, JHEP {\bf 03} 101 (2022).
\bibitem{8}T.T. Wang, S.M. Zhao, X.X. Dong, et al., $Lepton~flavor~violating~decays~\tau\rightarrow~pl~in~the
\nonumber\\~U(1)_XSSM~model$, JHEP {\bf 04} 122 (2022).
\bibitem{9}B. Yan, S.M. Zhao and T.F. Feng, $Electric~dipole~moments~of~neutron~and~heavy~quarks~c,
 \nonumber\\t~in~CP ~violating~U(1)_XSSM$, Nucl. Phys. B, {\bf 975} 115671 (2022).
\bibitem{23}J. McDonald. $Gauge~singlet~scalars~as~cold~dark~matter[J]$. Phys. Rev. D. 50: 3637 (1994).
\bibitem{24}G. Blanger and F. Boudjema. $micrOMEGAs4.1:~two~dark~matter~candidates[J]$. Comput. Phys. Commun. 192: 322 (2015).
\bibitem{25}W.~Chao, H.K.~Guo, H.L.~Li. $Electron~flavored~dark~matter[J]$. Phys. Lett. B, 782:517(2018).
\bibitem{26}S. Gopalakrishna, et al., $Right-handed~sneutrinos~as~nonthermal~dark~matter$, JCAP {\bf 0605} 005 (2006).
\bibitem{27}C. P. Burgess and M. Pospeloy. $The~Minimal~model~of~nonbaryonic~dark~matter:~A~Singlet~scalar[J]$. Nucl.Phys. B. 619: 709 (2001).
\bibitem{28}X.G. He T. Li, X. Q. Li. $Constraints~on~Scalar~Dark~Matter~from~Direct~Experimental~Searches[J]$. Phys. Rev. D. 79: 023521 (2009).
\bibitem{29}J. L. Feng and K. T. Matchev. $Neutralino~dark~matter~in~focus~point~supersymmetry[J]$. Phys. Lett. B. 482: 388 (2000).
\bibitem{30}Wei Chao, Yong-chao Zhang. $Majorana~Dark~matter~with~B+L~gauge~symmetry[J]$. JHEP. 1704: 034 (2017).
\bibitem{31}H. Baer, C. Balazs, A. Belyaev, $Neutralino~relic~density~in~minimal~supergravity~with~co-annihilations$, JHEP. {\bf 03} 042 (2002).
\bibitem{32}J. Edsj\"{o} and P. Gondolo, $Neutralino~relic~density~including~coannihilations$, Phys. Rev. D. {\bf 56} 1879 (1997).
\bibitem{34}G. Jungman, M. Kamionkowski, K. Griest, $Supersymmetric~dark~matter$, Phys. Rep. {\bf 267} 195 (1996).
\bibitem{33}G. Bertone, D. Hooper, J. Silk, $Particle~dark~matter:~Evidence,~candidates~and~constraints$, Phys. Rep. {\bf 405}  279 (2005).
\bibitem{35}M. Freytsis, Z. Ligeti, $On~dark~matter~models~with~uniquely~spin-dependent~detection~possibilities$, Phys. Rev. D, {\bf 83} 115009 (2011).
\bibitem{36}T. Bringmann, J. Edsjo, P. Gondolo, et al., $DarkSUSY~6:~An~Advanced~Tool~to~Compute~Dark
\nonumber\\~Matter~Properties~Numerically$, JCAP 1807 033 (2018).
\bibitem{37}W. Chao, $Direct~detections~of~Majorana~dark~matter~in~vector~portal$, JHEP 1911 013 (2019).



\end{thebibliography}
\end{document}